\documentclass[preprint,aps,superscriptaddress,nofootinbib]{revtex4}
\usepackage{amsmath}
\usepackage{graphicx}
\usepackage{color}
\usepackage{slashed}

\begin{document}

\title{The small and large $D$ limit of Parikh-Wilczek tunneling model for Hawking radiation}
\author{June-Yu Wei}\thanks{%
E-mail: weijuneyu@gmail.com}
\affiliation{Department of Physics and Center for High Energy Physics, Chung Yuan Christian University, Chung Li City, Taiwan}
\author{Wen-Yu Wen}\thanks{%
E-mail: steve.wen@gmail.com}
\affiliation{Department of Physics and Center for High Energy Physics, Chung Yuan Christian University, Chung Li City, Taiwan}
\affiliation{Leung Center for Cosmology and Particle Astrophysics\\
National Taiwan University, Taipei 106, Taiwan}

\begin{abstract}
In this note, we study both the small and large dimension $D$ limit of the tunneling model of Hwaking radiation proposed by Parikh and Wilczek\cite{Parikh:1999mf}.  We confirm that the result $\Gamma \sim e^{\Delta S}$ is still valid for arbitrary $D>3$.   The  sensible large $D$ limit is given by $\sqrt{D} \ll r_0$ in order to have nonzero radiation.  On the other hand, the sensible small $D$ limit is given by taking $D=3+\epsilon$ as a continuous parameter.  We also explicitly show the leading order correction to the thermal radiation and discuss its connection to the two-dimensional dilaton gravity.
\end{abstract}

\maketitle



\section{Introduction}
It was argued and shown in \cite{Emparan:2013moa} that in the limit $N\to \infty$, General Relativity dramatically simplifies and $1/N$ could serve as a useful expansion parameter.  Dynamics of a black hole with the horizon $r_0$ is controlled by two relevant small scales: $r_0/\sqrt{D}$ due to horizon area and $r_0/D$ form the strong localization of gravitational potential.  
In this paper, we would like to confirm prediction from the tunneling model of Hawking radiation proposed by Parikh and Wilczek and study its nonthermal correction in the limits of large and small $D$.  We start with a brief review of the $D$-dimensional black hole metric and its thermal quantities.  A $D$-dimensional Schwarzschild black hole metric is given in the Schwarzschild-Tangherlini form\cite{Tangherlini:1963}:
\begin{eqnarray}\label{ss_metric}
ds^2 =&& -f(r)dt_s^2 + f(r)^{-1}dr^2+r^2d\Omega_{D-2}^2,\nonumber\\
&& f(r) = 1-(\frac{r_0}{r})^{D-3}
\end{eqnarray} 
where $r=r_0$ is the horizon such that $r_0^{D-3}=\frac{16\pi}{(D-2)\Omega_{D-2}}M$.  The volume of unit sphere $\Omega_{D-2}=\frac{2\pi^{(D-1)/2}}{\Gamma(\frac{D-1}{2})}$.  The Hawking temperature is given by 
\begin{equation}\label{Hawking_T}
T_H=\frac{D-3}{4\pi r_0}.
\end{equation}
The apparent singularity at the horizon can be removed by a coordinate transformation of the Painlev\`{e}-type:
\begin{equation}
t = t_s + r_0^{(D-3)/2}\int{\frac{dr}{\sqrt{r^{D-3}-r_0^{D-3}}}}, 
\end{equation}
which brings into the metric suitable for our purpose:
\begin{equation}
ds^2 = -(1-(r_0/r)^{D-3})dt^2 + 2(\frac{r_0}{r})^{(D-3)/2}dt dr + dr^2 + r^2 d\Omega_{D-2}^2. 
\end{equation}
We remark that the tunneling process across the horizon happens in this stationary vacuum, instead of static one given by metric (\ref{ss_metric}).

\section{Tunneling model in $D$ dimensions}
In this section, we apply the Parikh-Wilczek tunneling model\cite{Parikh:1999mf} to $D$-dimensional black hole and obtain the general expression for tunneling rate in $D$ dimensions.  We regard radiation as a shell of energy $\omega$ moving in the geodesics of  a spacetime with $M$ replaced by $M-\omega$ due to back reaction, where the radial null geodesics are given by 
\begin{equation}
\dot{r} \equiv \frac{dr}{dt} = \pm 1 - \sqrt{(\frac{r_\omega}{r})^{D-3}}.
\end{equation}
$r_\omega$ denotes the horizon after radiation, such that $r_\omega^{D-3} = \frac{16\pi}{(D-2)\Omega_{D-2}}(M-\omega)$.  The WKB approximation states that the emission rate $\Gamma \sim e^{-2 Im S}$, where the imaginary part of action reads
\begin{equation}\label{integral}
Im S = Im \int_{r_{in}}^{r_{out}}{p_r dr} = Im \int_M^{M-\omega}\int_{r_0}^{r_\omega}{\frac{dH}{\dot{r}}dr},
\end{equation}
for Hamiltonian $H = M-\omega^\prime$.  The half line contour integral in the complexified $r$-plane picks up a residue at $r=r_{\omega^\prime}$, that is 
\begin{equation}\label{contour}
\int_{r_0}^{r_\omega}{\frac{dr}{\dot{r}}}=\frac{2\pi i}{D-3}\big(\frac{16\pi}{(D-2)\Omega_{D-2}}(M-\omega^\prime)\big)^{1/(D-3)}.
\end{equation}
After substituting this result into integral (\ref{integral}), one obtains
\begin{equation}\label{entropy_change}
-2 Im S = \frac{4\pi}{D-2}\big(\frac{16\pi}{(D-2)\Omega_{D-2}}\big)^{\frac{1}{D-3}}M^{\frac{D-2}{D-3}}\big((1-\frac{\omega}{M})^{\frac{D-2}{D-3}}-1\big) = \Delta S_{BH}, 
\end{equation}
which is exact the change of Bekenstein-Hawking entropy after radiation of energy $\omega$ for a $D$-dimensional Schwarzschild black hole.  This explicit calculation illustrates that the Parikh-Wilczek tunneling model of Hawking radiation in arbitrary dimension $D>3$ predicts the same result, that is $\Gamma \sim e^{\Delta S_{BH}}$.

\section{Large $D$ correction to the emission rate}
In this section, we take large $D$ limit of emission rate to see its correction to thermal spectrum.
For $\omega \ll M$, we can expand
\begin{equation}\label{expansion}
\Delta S \simeq -\frac{4\pi}{D-3}r_0 \omega + \frac{4\pi}{(D-3)^2}r_0\frac{\omega^2}{M}+\cdots
\end{equation}
The first term is nothing but Hawking's thermal radiation.  We see that naively sending $D\to \infty$ corresponds to a black hole at infinite temperature.  If at the same time keeping $r_0/D$ finite, it corresponds to that at finite temperature.  Both limits give rise to a trivial emission rate thanks to vanishing first term in (\ref{expansion}) for sufficiently small $\omega$.  On the other hand, the sensible limit which keeps the first term finite while $\omega \ll M$ is given by that  $r_0/\sqrt{D} \gg 1$.  It was pointed out in \cite{Emparan:2013moa} that the length scales like $r_0/\sqrt{D}$ in $D$ dimensions, so this nontrivial limit asks for a black hole of sizable area to ensure a significant radiation.  The emission rate after receiving nonthermal correction from the second term in (\ref{expansion}) reads $\Gamma = e^{(\sqrt{D}/r_0)^D}\Gamma_0$, where the usual Hawking thermal radiation $\Gamma_0 \equiv e^{-\omega/{T_H}}$ and $\omega/T_H \sim {\cal O}(1)$ is assumed.  Practically speaking, for a ratio $r_0/ \sqrt{D}\simeq {\cal O}(100)$, the correction only appears at the eighth decimal place for $D=4$, and neglectable for any large $D$.

\section{Small $D$ correction to the emission rate}
It is also interesting to investigate (\ref{expansion}) for small dimensions, say $D \to 3^+$\cite{Asnin:2007rw}.  To do that, we let $D=3+\epsilon$ and then send $\epsilon \to 0$.  This limit gives rise to zero temperature and no radiation.  For nonvanishing thermal radiation, one needs  that $r_0 \omega \simeq \epsilon$ for a finite first term in (\ref{expansion}), while importance of the nonthermal correction could vary with different scaling.  For instance, if one further scales $\omega/M \sim {\cal O}(\epsilon)$,  the correction term in (\ref{expansion}) would have the same order as the first one.

\section{Discussion}
We have demonstrated that the tunneling model of Hawking radiation predicts the same result $\Gamma \sim e^{\Delta S}$ in arbitrary dimension $D$.  In particular, one can fix the scale $r_0/D$ for nonvanishing Hawking temperature, or $r_0/\sqrt{D}$ for nonzero Hawking radiation.  The nonthermal correction is also obtained in terms of $D$.  Several comments are in order: 

Firstly, a different way of taking large $D$ limit of black hole was studied in the dimensionally continued gravity\cite{Giribet:2013wia}, where the function $f(r)$ in the metric (\ref{ss_metric}) is replaced by $f(r)=1+\frac{r^2}{L^2}-(\frac{r_0}{r})^\frac{2(D-D_c)}{(D_c-1)}$, for some critical dimension $D_c \ge 3$.  In particular, in the limit $D,L\to \infty$ but keeping$D/D_c$ fixed, the solution becomes asymptotically flat and temperature is finite.  Due to the higher-curvature terms in the action, the Bekenstein-Hawking area law is disobeyed and $S \propto A^{\frac{D-D_c}{D}}$.  Nevertheless, we still find that the result $\Gamma \sim e^{\Delta S}$ is valid.  Our observation is that the Parikh-Wilczek tunneling model predicts the same result as long as the first law of thermodynamics is satisfied\footnote{In this particular limit, the black hole mass $M\sim \frac{r_0^{D-D_c}}{2G}$, Hawking temperature $T\sim \frac{\hbar(D-D_c)}{2\pi D_c r_0}$ and Bekenstein-Hawking entropy $S\sim \frac{\pi D_c r_0^{D-D_c}}{\hbar G (D-D_c)}$\cite{Giribet:2013wia}.  It is straightforward to check that the first law of thermodynamics $dM=TdS$ holds.}.  This reflects the fact that the tunneling model respects conservation of energy, which is the very spirit of the first law.   Observing our result that emission rate for tunneling model takes a universal form $\Gamma \sim e^{\Delta S}$ for arbitray dimension $D > 3$ and extended theories of gravity which respects energy conservation, one tends to believe that the concept of spacetime may be irrelevant in some parts of formulation of quantum gravity, as suggested in the \cite{Braunstein:2011gz}.

Secondly, the Parikh-Wilczek model, as a WKB approximation, can receive further quantum correction.  One example was provided by \cite{Banerjee:2008ry} for one-loop correction to surface gravity.  As a result, the Bekenstein-Hawking area law receives a logarithmic correction, that is $S=4\pi M^2 -4\pi \alpha \ln{(1+\frac{M^2}{\alpha})}$, where $\alpha$ is related to the trace anomaly.  The change of entopy is modified as\cite{Banerjee:2008ry}:
\begin{eqnarray}
\Delta S &=& -8\pi M\omega + 4\pi \omega^2 + 4\pi \alpha \ln{\frac{\alpha+M^2}{\alpha+(M-\omega)^2}}\nonumber\\
&\simeq& -8\pi M\omega +4\pi \omega^2 + \frac{2M}{M^2+\alpha}\omega + \cdots,
\end{eqnarray}
where approximation in the second line shows an expansion for small $\omega$.   If one considers the same limit adopted in the section III and IV for finite emission, that is $\omega/T_H \sim {\cal O}(1)$ or equivalently $\omega M \sim {\cal O}(1)$, then the logarithmic correction (last term on second line) could be comparable to the nonthermal term for $\alpha \ll M^2$.  However, one expects that $\alpha$ also grows like $D$ and could compete with the $M^2$ term, which grows like $r_0^{2D}/D^D\gg 1$.  Then the logarithmic correction could be ignored if $r_0$ is chosen such that $\alpha \gg M^2$.

At last, we would like to comment on its connection to the dilaton gravity in two dimensions\cite{Grumiller:2002nm}.   Following the discussion in \cite{Grumiller:2002nm,Carlip:2011uc}, the Euclidean $D$-dimensional Schwarzschild black hole solution can be dimensionally reduced to a dilatonic black hole in two dimensions with the action:
\begin{eqnarray}
I = -\frac{1}{2G_2} \int{d^2x} \sqrt{g} \big[ XR-U(X)(\partial X)^2 -2V(X) \big],
\end{eqnarray}
where the dilaton field $X=X(r)$ for a static solution.  Several relevant functions and potentials are defined by
\begin{eqnarray}
&&U(X)=-\frac{1}{X}\frac{D-3}{D-2},\nonumber\\
&&Q(X) = \int^X {dX^\prime U(X^{\prime})},\nonumber\\
&&V(X) = \lambda e^{-2Q(X)} U(X),\nonumber\\
&&w(X)=-2\int^X{dX^\prime e^{Q(X^\prime)}V(X^\prime) }.
\end{eqnarray}
The Hawking temperature then is given by 
\begin{equation}
T_H = \frac{w^{\prime}(X)}{4\pi}\big|_{X=X_h}.
\end{equation}
Here without loss of generality, we normalized the coefficient of proportionality $\lambda$ such that $X=X_h=1$ at the horizon.   For large $D$, one can reproduce the Hawking temperature in (\ref{Hawking_T}) for the following assignment of functions:
\begin{eqnarray}
&&U(X) \to -\frac{1}{X}, \qquad Q = -\ln{X},\nonumber\\
&&V(X) = -\lambda X, \qquad w = 2\lambda X,
\end{eqnarray}
and $\lambda = \frac{D-3}{2r_0}$.  As to $D=3+\epsilon$ with small $\epsilon$, one obtains:
\begin{eqnarray}
&&U(X) \to -\frac{\epsilon}{X}, \qquad Q = -\epsilon\ln{X},\nonumber\\
&&V(X) = -\lambda \epsilon X^{2\epsilon-1}, \qquad w = 2\lambda X^{\epsilon},
\end{eqnarray}
and the choice $\lambda=\frac{1}{2r_0}$.  The limit $\epsilon \to 0$ gives rise to trivial potentials and constant $w=2\lambda$, implying zero Hawking temperature and no radiation.  This result agrees with the observation in the section IV.  It is worth seeking a model of varying effective dimensions to match the nonthermal radiation in the Parikh-Wilczek model with or without one-loop correction.  We will leave this for the future study.

\begin{acknowledgments}
We are grateful to the help of Ji-Hsien Dai at the early stage of this work.  This work is supported in parts by the Taiwan's Ministry of Science and Technology (grant No. 102-2112-M-033-003-MY4) and the National Center for Theoretical Science. 
\end{acknowledgments}


\end{document}